\documentclass[9pt,twocolumn,twoside]{opticajnl}
\journal{opticajournal} 

\setboolean{shortarticle}{true}

\usepackage{lineno}

\title{Resolution-Agnostic Lensless Imaging via Fourier Neural Operators}

\author[1]{Kerem Ekec}
\author[1,2,*]{U\u{g}ur Te\u{g}in}

\affil[1]{Ko\c{c} University, Department of Biomedical Sciences and Engineering, \.{I}stanbul, T\"{u}rkiye}
\affil[2]{Ko\c{c} University, Department of Electrical and Electronics Engineering, \.{I}stanbul, T\"{u}rkiye}
\affil[*]{utegin@ku.edu.tr}

\begin{abstract}
Lensless cameras based on thin diffusers offer a compact alternative to conventional refractive imaging  but rely on computational reconstruction, since the diffuser's point spread function (PSF) globally  multiplexes every scene point across the sensor. Here, we report a Fourier Neural Operator (FNO)  framework for this reconstruction task. Because a linear shift-invariant forward model reduces to  a pointwise multiplication in Fourier space, the spectral-domain kernel of an FNO layer is  structurally aligned with the DiffuserCam inverse problem. Using a compact DiffuserCam prototype  and a 25,000-image natural-scene dataset, our FNO improves upon a U-Net baseline of comparable  parameter count by $2.14$~dB in PSNR and $0.11$ in SSIM. The same FNO, trained exclusively at  $128 \times 128$, reconstructs $256 \times 256$ and $512 \times 512$ measurements with less  than $1$~dB loss in PSNR and no retraining, demonstrating resolution-agnostic inference.  The framework is directly applicable to other lensless modalities with global PSFs, such as  multimode-fiber endoscopy.\end{abstract}

\setboolean{displaycopyright}{false} 

\begin{document}

\maketitle

Lensless imaging has emerged as a compelling route toward ultra-compact and lightweight computational cameras, in which a thin optical element---a diffuser, coded amplitude mask, or multimode fiber---replaces conventional refractive optics in front of the sensor \cite{aydogan, boominathan_recent_2022, tegin1, tegin2, Ioannis}. Rather than focusing light, the scattering element globally multiplexes scene information across the sensor plane through its point spread function (PSF), so that image formation is delegated entirely to a computational reconstruction step. Assuming linear, shift-invariant scattering, the raw measurement can be written as

\begin{equation}
    y = h * x + \eta,
\end{equation}

\noindent
where $x$ is the scene, $h$ is the PSF of the diffuser, $*$ denotes convolution, and $\eta$ is measurement noise. In a DiffuserCam configuration \cite{antipa_diffusercam_2018}, each scene point contributes to a large fraction of the sensor area. This inherent multiplexing enables lensless cameras to capture higher-dimensional signals such as 3D volumes or video from a single 2D frame, but it also forces the inverse problem to be solved over the full measurement.

Classical reconstructions address this ill-posed problem by iterative optimization with hand-crafted priors such as total variation or sparsity \cite{boyd}, which is effective but computationally expensive and sensitive to tuning. Deep learning has since become a powerful alternative across computational imaging, including super-resolution, holography, and phase retrieval \cite{phase,liu_super}. For diffuser-based lensless systems in particular, convolutional neural networks have been trained end-to-end to map raw sensor measurements to reconstructions \cite{barba}, and hybrid architectures that combine a model-based inversion with a CNN post-processor---notably FlatNet \cite{flat} and the unrolled Le-ADMM-Net \cite{monakhova}---have emerged as strong benchmarks. Despite their success, these networks are built from localized convolutional kernels and operate on fixed pixel grids. Neither property is naturally aligned with the global PSF of the diffuser, and generalization across different imaging resolutions typically requires retraining.

Neural operators (NOs) offer a different formulation: instead of learning mappings between finite-dimensional arrays, they learn mappings between function spaces and are, in principle, agnostic to the underlying discretization \cite{duruisseaux2025fourier}. Originally introduced to solve partial differential equations, NOs have since been applied to inverse problems in imaging \cite{csmri} and to super-resolution \cite{srno}. Of particular relevance here is the Fourier Neural Operator (FNO), whose core layer parameterizes a global integral kernel directly in the spectral domain \cite{li2021fourier}. Because a linear shift-invariant system is a pointwise multiplication in Fourier space, this spectral-domain kernel is structurally aligned with the DiffuserCam forward model of Eq.~(1). In addition, the learned spectral weights act on continuous function representations, so a trained FNO can be evaluated at a different spatial discretization without architectural modification.

Here, we report an FNO-based reconstruction framework for lensless diffuser imaging. We build a compact DiffuserCam prototype, collect a paired natural-scene dataset from the MIR-Flickr collection, and train an FNO and a U-Net baseline with matched parameter count under identical optimization settings. At the training resolution, the FNO improves the U-Net baseline by 2.14~dB in PSNR and 0.11 in SSIM, while requiring less per-epoch training time and less GPU memory. More importantly, the same FNO---trained exclusively at $128 \times 128$---reconstructs $256 \times 256$ and $512 \times 512$ measurements with less than 1~dB loss in PSNR and no retraining, even though these higher-resolution inputs contain spatial frequencies above the training Nyquist limit. These results suggest that the structure of the reconstruction problem, a global convolution naturally inverted in the spectral domain, is a useful guide for the choice of network architecture and that neural operators provide a practical path toward resolution-agnostic lensless imaging.

\begin{figure}[t!]
\centering
\includegraphics[width=\linewidth]{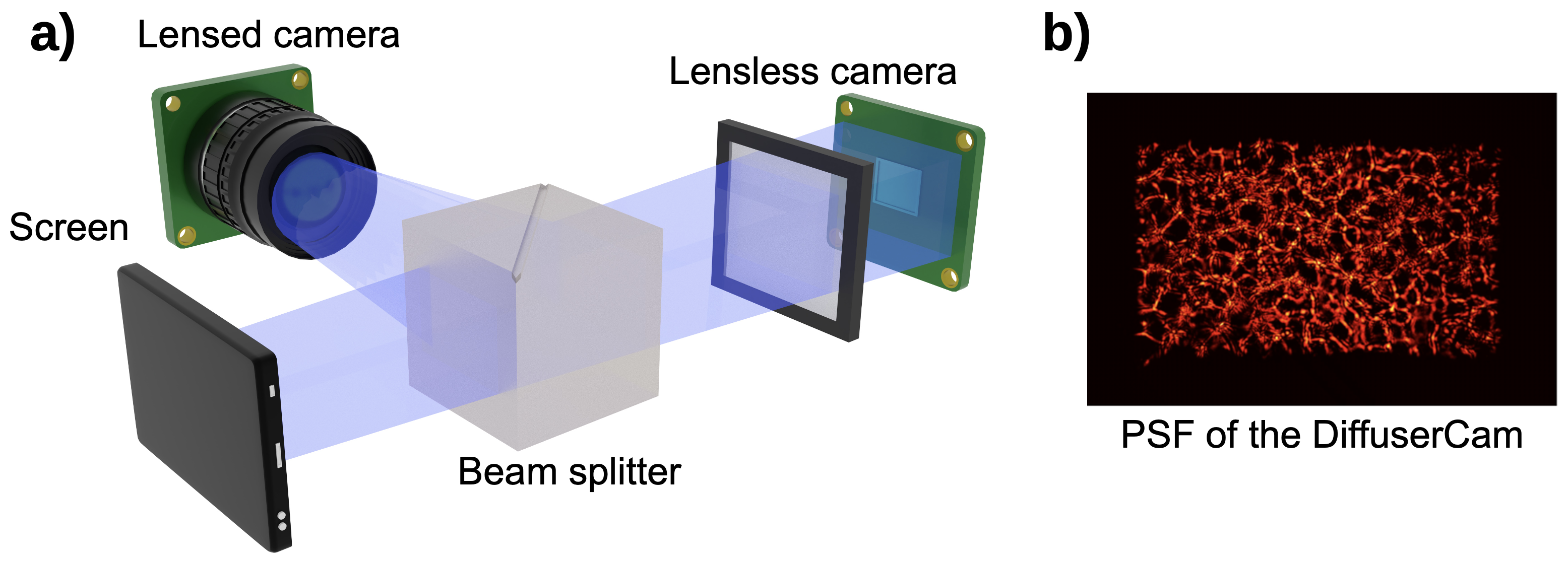}
\caption{Experimental setup of the DiffuserCam. Before data acquisition, the system is calibrated and aligned following the procedure in \cite{antipa_diffusercam_2018}. The measured PSF has sharp features and covers most of the sensor area, giving a global mapping from the scene to the measurement.}
\label{setup}
\end{figure}

To collect the paired dataset required for training, we constructed a DiffuserCam system with a dual-aperture acquisition scheme. A tablet screen is used as the scene source, and its emitted field is split by a 50:50 beam splitter into two optical paths. In the first path, a conventional lensed camera captures a focused ground-truth image. In the second path, a bare sensor of the same model is placed behind a thin diffuser (double-sided Scotch tape) at approximately 3~mm from the sensor, capturing the diffused measurement. The two cameras are spatially aligned so that their fields of view cover the same region of the tablet screen. A software trigger synchronizes the tablet display and both cameras, enabling automated sequential capture of the full dataset. The system layout is shown in Fig.~\ref{setup}.

We used the MIR-Flickr-25k dataset \cite{huiskes08} as the source of natural-scene images in order to test reconstruction on complex and diverse content. A total of 25{,}000 images were sequentially displayed on the tablet and simultaneously captured by both cameras at the native sensor resolution of $1240 \times 720$~pixels. Each captured pair was then resized to $512 \times 512$ to serve as the highest-resolution ground truth and corresponding measurement. The dataset was split into 23{,}000 training pairs, 1{,}000 validation pairs, and 1{,}000 test pairs. For display, the central region of the reconstructed image is cropped to remove the black border inherited from the ground-truth image.

To investigate the resolution-agnostic reconstruction capability of the FNO, we constructed a multi-resolution evaluation pipeline. The $512 \times 512$ image pairs were downsampled by factors of 4 and 2 using bilinear interpolation, producing paired datasets at $128 \times 128$ and $256 \times 256$ resolution, respectively. The $128 \times 128$ pairs were used exclusively for training. The $256 \times 256$ and $512 \times 512$ pairs were reserved for evaluating the trained network on discretizations unseen during training, which we refer to as resolution-agnostic or discretization-invariant inference. We emphasize that the underlying physical system is unchanged across these evaluations: the diffuser, the sensor, and the measured PSF $h$ in Eq.~(1) are identical, and only the input discretization of the measurement $y$ varies. Because the FNO operates on continuous function representations rather than a fixed pixel grid, the same trained network can ingest these differently-sampled measurements without any architectural modification. Higher-resolution inputs additionally carry spatial-frequency content above the Nyquist limit of the $128 \times 128$ training data --- fine PSF structure that is aliased away by the $4\times$ downsampling --- so this evaluation also probes the robustness of the network to signal components that are, by construction, absent from the training distribution.

\begin{figure}[b!]
\centering
\includegraphics[width=\linewidth]{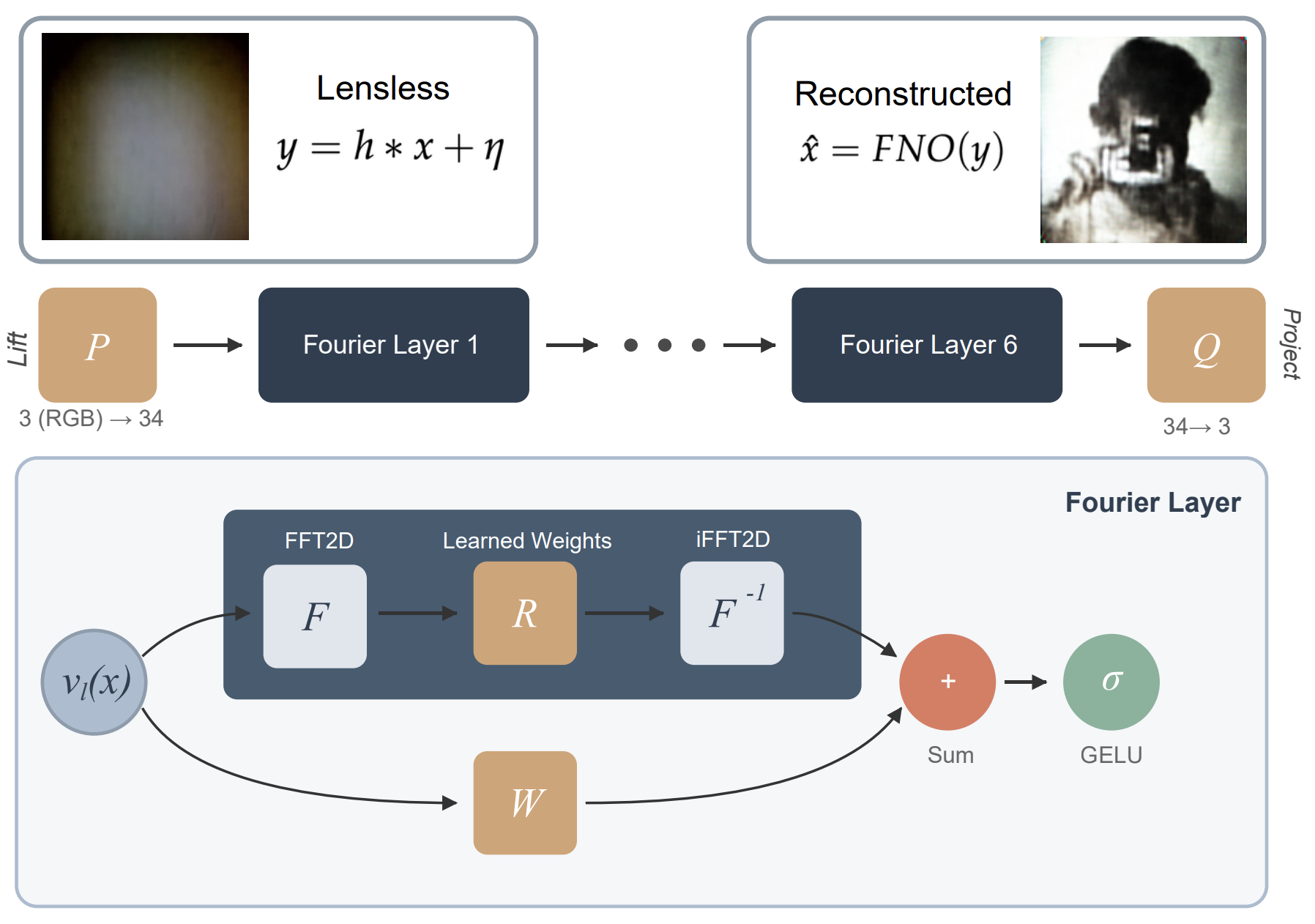}
\caption{FNO architecture. A raw lensless measurement is processed through six Fourier layers, each performing learned spectral filtering via a 2D FFT, complex-valued weight multiplication $R$, and inverse 2D FFT, with a local $1\times1$ convolution bypass. Because the filtering operates in Fourier space, the network naturally captures global dependencies and is agnostic to the input discretization.}
\label{network}
\end{figure}

As a comparative baseline, we trained a U-Net following the architecture in \cite{Unet}. The encoder–decoder has six levels with three downsampling and three upsampling stages and skip connections at each scale, totalling 7{,}785{,}859 learnable parameters. The input is the three-channel (RGB) diffused measurement and the output is the three-channel reconstructed image.

The FNO architecture consists of an input lifting layer, six sequential Fourier layers, and a projection layer. The three-channel RGB input is first lifted to a 34-dimensional channel space via a pointwise linear transformation. Each Fourier layer implements
\begin{equation}
v_{l+1}(\mathbf{r}) = \sigma \left( W_l v_l(\mathbf{r}) + \mathcal{F}^{-1} \left( R_l \cdot \mathcal{F} [v_l] \right) \right),
\end{equation}
where $\mathcal{F}$ and $\mathcal{F}^{-1}$ denote the 2D Fast Fourier Transform and its inverse, $R_l(\mathbf{k})$ is a learnable complex-valued spectral weight tensor, $W_l$ is a pointwise linear transformation acting as a local residual path, and $\sigma$ is a GELU nonlinearity. The multiplication $R_l \cdot \mathcal{F}[v_l]$ in Fourier space corresponds to a convolution with a kernel of global spatial support, so each output point depends on the entire input field. We retain 24 modes in each spatial dimension ($24 \times 24$ modes in total). For $128 \times 128$ inputs, the maximum representable number of modes is 64 (the Nyquist limit), so 24 modes capture roughly the lowest 37\% of the spatial-frequency spectrum. This choice was driven primarily by the need to keep the FNO's total parameter count (8.0~M) comparable to that of the U-Net baseline (7.8~M), ensuring a fair architectural comparison. In preliminary experiments, we also trained FNO variants with a larger number of retained modes; reconstruction quality continued to improve, but at the cost of a proportional growth in parameter count and memory footprint that would have broken parity with the U-Net. We therefore fixed the mode count at 24 throughout the rest of this study and leave a systematic mode-count sweep to future work.

After the final Fourier layer, the 34-channel representation is projected back to three channels via a pointwise linear layer. The total number of learnable parameters is 8{,}002{,}483. Because the Fourier modes are defined over continuous frequencies rather than pixel grids, the trained FNO can accept inputs at any spatial resolution without architectural modification. The architecture is shown in Fig.~\ref{network}.

Both networks were implemented in PyTorch and trained on a single NVIDIA A100 GPU. We used the AdamW optimizer with an initial learning rate of $1 \times 10^{-3}$, a weight decay of $1 \times 10^{-4}$, and a batch size of 32. A cosine annealing schedule decayed the learning rate to a minimum of $1 \times 10^{-6}$ over the course of training. Both networks were trained for 50 epochs, and the FNO was observed to converge in fewer epochs than the U-Net.

The training loss combines a joint SSIM-$L_1$ term with a VGG-based perceptual term,
\begin{equation}
\mathcal{L} = \alpha\,\mathcal{L}_{\mathrm{SSIM}\text{-}L_1} + \beta\,\mathcal{L}_{\mathrm{VGG}},
\end{equation}
where $\mathcal{L}_{\mathrm{SSIM}\text{-}L_1}$ is the standard joint structural-similarity / $L_1$ pixel fidelity term and $\mathcal{L}_{\mathrm{VGG}}$ is a perceptual loss computed on intermediate features of a pretrained VGG-16 network. The SSIM-$L_1$ term enforces pixel-level accuracy and local structure, while the VGG term penalizes deviations in higher-level perceptual features, which we found to sharpen fine details in the reconstructions. The weights $\alpha$ and $\beta$ were selected on the validation set. The same loss, optimizer, and hyperparameters were used for both networks to ensure a fair comparison. To evaluate the resolution-agnostic capability of the FNO, the network trained exclusively at $128 \times 128$ was tested on higher-resolution inputs ($256 \times 256$ and $512 \times 512$) without retraining. Reconstruction quality was assessed via PSNR and SSIM against the corresponding ground truth at each evaluation resolution.

\begin{table}[htbp]
\centering
\caption{\textbf{Comparison of Average Performance Metrics between U-Net and FNO}$^a$}
\begin{tabular}{lcccc} 
\hline
Method & Parameters & PSNR (dB) & SSIM & Time/Epoch \\ \hline
U-Net  & 7,785,859  & 20.18     & 0.65 & 102s        \\
FNO    & 8,002,483  & 22.32     & 0.76 & 93s       \\ \hline
\end{tabular}
\label{tab:results-comparison}

\smallskip
\small
$^a$Results are evaluated on the held-out test set at $128 \times 128$ resolution.
\end{table}

\begin{figure}[ht]
\centering
\includegraphics[width=\linewidth]{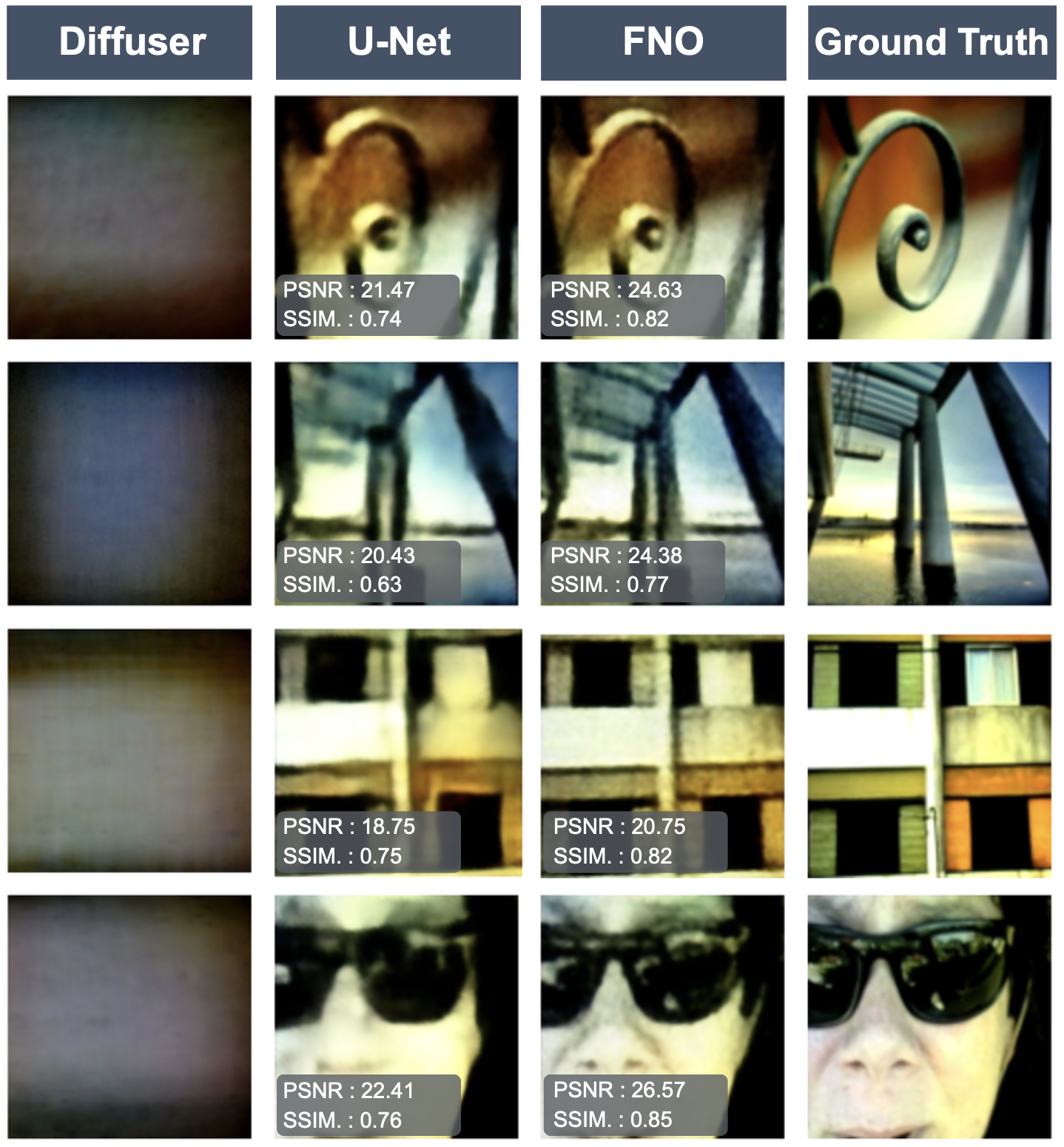}
\caption{Comparison of image reconstruction with the FNO and the U-Net. The FNO yields higher PSNR and SSIM across test images of varying complexity.}
\label{fig:Uvsfno}
\end{figure}

Both networks were trained on the $128 \times 128$ dataset under identical conditions (same loss, optimizer, batch size, and number of epochs) to ensure a fair comparison. We chose the U-Net as our baseline because it represents a well-established data-driven approach to lensless image reconstruction; by keeping both methods purely learning-based, without explicit incorporation of forward-model knowledge, we isolate the effect of architectural design on reconstruction performance. Table~\ref{tab:results-comparison} reports PSNR and SSIM averaged over the 1{,}000-image test set. The FNO achieves a PSNR of 22.32~dB and an SSIM of 0.76, compared with 20.18~dB and 0.65 for the U-Net, corresponding to an improvement of 2.14~dB in PSNR and 0.11 in SSIM. Representative reconstructions from the test set are shown in Fig.~\ref{fig:Uvsfno}.

We attribute this improvement to the FNO's global spectral-domain processing. By the convolution theorem, the forward model in Eq.~(1) becomes a pointwise multiplication $\hat{y}(\mathbf{k}) = \hat{h}(\mathbf{k})\,\hat{x}(\mathbf{k}) + \hat{\eta}(\mathbf{k})$ in Fourier space, so inverting the diffuser blur is, at its core, a spectral operation. An FNO layer learns precisely this class of operation end-to-end: its spectral weight tensor $R_l(\mathbf{k})$ acts pointwise in frequency and therefore corresponds to a convolution with a kernel of global spatial support in the physical domain. Reconstruction of each pixel can thus draw on information from the entire measurement in a single layer. The U-Net, by contrast, must expand its effective receptive field indirectly through hierarchical pooling and skip connections, which appears less well matched to a forward model whose PSF already covers most of the sensor area.

The FNO also shows a modest training-cost advantage. The per-epoch training time was 93~s for the FNO and 102~s for the U-Net, and the FNO and U-Net used 3.9~GB and 5.2~GB of VRAM during training, respectively.

\begin{table}[htbp]
\centering
\caption{\bf Resolution-Agnostic Performance of FNO$^\textit{a}$}
\begin{tabular}{ccccc}
\hline
Method & Evaluated at & PSNR (dB) & SSIM  \\
\hline
FNO  & $128 \times 128$ & 22.32 & 0.76 \\
FNO  & $256 \times 256$ & 21.54 & 0.73 \\
FNO  & $512 \times 512$ & 21.35 & 0.72 \\
\hline
\end{tabular}
\label{tab:fno-resolution-invariance}

\smallskip
$^\textit{a}$Note that the model was trained exclusively on $128 \times 128$ data.
\end{table}

A key property of neural operators is their ability to generalize across discretizations, since they learn mappings between continuous function spaces rather than between fixed pixel grids. To illustrate this in the present setting, we evaluated the FNO trained exclusively at $128 \times 128$ on measurements at $256 \times 256$ and $512 \times 512$, i.e., resolutions unseen during training. Increasing the resolution corresponds to a denser discretization of the same physical field of view: the underlying scene, diffuser, and PSF are all held fixed, and only the sampling density of the measurement changes. This is a non-trivial test in two respects. First, the higher-resolution measurements contain fine PSF structure --- spatial-frequency components above the Nyquist limit of the $128 \times 128$ training data --- that are aliased away by the $4\times$ downsampling and are therefore absent from the training distribution by construction. Second, a standard CNN trained on $128 \times 128$ inputs cannot accept a $512 \times 512$ input at all without either retraining or resorting to tiling or input downsampling, whereas the FNO accepts it natively. Table~\ref{tab:fno-resolution-invariance} summarizes the quantitative results. The ground truth at each evaluation resolution is the corresponding downsampled version of the $512 \times 512$ reference.

\begin{figure}[t!]
\centering
\includegraphics[width=\linewidth]{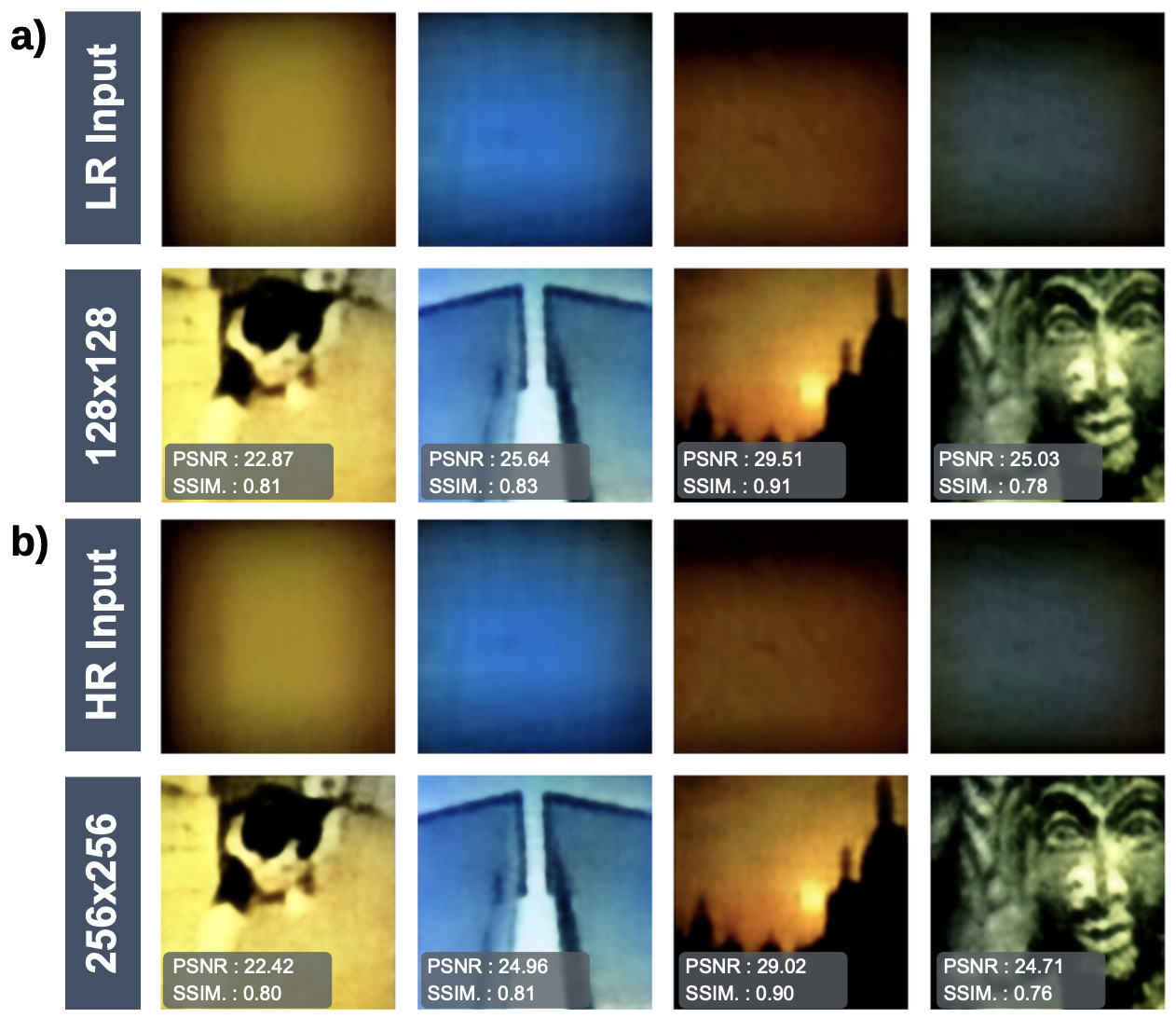}
\caption{Resolution-agnostic inference with the FNO. (a) The network is trained exclusively on $128\times128$ images. (b) At test time, the same network is evaluated on $256\times256$ measurements that contain spatial frequencies absent from the training distribution. The reconstruction quality is largely preserved, consistent with generalization of the learned operator across discretizations.}
\label{fig:supres}
\end{figure}

Several broader implications and limitations of these results deserve comment. The architectural-match argument exploited here is not specific to DiffuserCam: any imaging modality whose forward model reduces to a global convolution between the scene and a system-wide PSF --- including computational microscopy through scattering media, multimode-fiber endoscopy, and single-pixel imaging --- should in principle benefit from a spectral-domain reconstruction of the kind an FNO layer naturally implements, and the same discretization-invariant property should carry over. The efficiency advantage observed in our experiments (3.9~GB vs.\ 5.2~GB of VRAM and 93~s vs.\ 102~s per training epoch) is a concrete practical benefit for deployment on memory-constrained hardware, where the ability to train at one resolution and infer at another with a single network is especially attractive for embedded lensless cameras and endoscopic front-ends. At the same time, the present study has clear limitations. The quantitative comparison is restricted to a purely learning-based U-Net baseline at matched parameter count; model-based hybrid methods such as FlatNet \cite{flat} and the unrolled Le-ADMM-Net \cite{monakhova} typically achieve higher reconstruction quality by injecting the measured PSF directly into the reconstruction pipeline, and a side-by-side comparison against these methods is left to future work. In addition, our training data consist of tablet-screen captures of natural images, which approximate but do not reproduce the depth-encoding behavior of real 3D scenes through a DiffuserCam.

In conclusion, we have demonstrated Fourier Neural Operators as a reconstruction backbone for lensless diffuser imaging, exploiting the structural match between the FNO's global spectral-domain kernel and the diffuser's global PSF. Our results indicate that an operator-theoretic view of the reconstruction network can simultaneously improve image quality at a fixed parameter budget and enable resolution-agnostic inference without retraining. We believe that this framework is of particular interest to lensless imaging modalities governed by a global PSF --- notably multimode-fiber endoscopy --- where a physics-informed FNO that incorporates the measured PSF directly into the operator is a natural next step.

\begin{backmatter}
\bmsection{Funding} This work was supported by the Scientific and Technological Research Council of Turkey (TUBITAK) under grant number 122C150.

\bmsection{Disclosures} The authors declare no conflicts of interest.

\bmsection{Data availability} The source code used in this study is available at Ref.~\cite{LPT2026LenslessFNO}.

\end{backmatter}

\bibliography{sample}

\bibliographyfullrefs{sample}


\ifthenelse{\equal{\journalref}{aop}}{%
\section*{Author Biographies}
\begingroup
\setlength\intextsep{0pt}
\begin{minipage}[t][6.3cm][t]{1.0\textwidth} 
  \begin{wrapfigure}{L}{0.25\textwidth}
    \includegraphics[width=0.25\textwidth]{john_smith.eps}
  \end{wrapfigure}
  \noindent
  {\bfseries John Smith} received his BSc (Mathematics) in 2000 from The University of Maryland. His research interests include lasers and optics.
\end{minipage}
\begin{minipage}{1.0\textwidth}
  \begin{wrapfigure}{L}{0.25\textwidth}
    \includegraphics[width=0.25\textwidth]{alice_smith.eps}
  \end{wrapfigure}
  \noindent
  {\bfseries Alice Smith} also received her BSc (Mathematics) in 2000 from The University of Maryland. Her research interests also include lasers and optics.
\end{minipage}
\endgroup
}{}

\end{document}